\documentclass[showpacs,aps,twocolumn]{revtex4}
\usepackage{epsfig}
\usepackage{graphicx}
\usepackage{amsmath,amssymb,amsfonts}
\usepackage{array}
\usepackage{url}
\usepackage{hyperref}
\usepackage{multirow}
\usepackage{float}
\usepackage{lineno}
\usepackage{xspace}
\usepackage{subfigure}
\usepackage[usenames,dvipsnames]{color}

\newcommand{\bef}{\begin{figure}}
\newcommand{\eef}{\end{figure}}
\newcommand{\bc}{\begin{center}}
\newcommand{\ec}{\end{center}}

\newcommand{\be}{\begin{equation}}
\newcommand{\ee}{\end{equation}}
\newcommand{\bea}{\begin{eqnarray}}
\newcommand{\eea}{\end{eqnarray}}

\def\ba{\begin{eqnarray}}
\def\ea{\end{eqnarray}}
\begin{document}
\title{Do Proton-Proton collisions at the LHC energies produce Droplets of Quark-Gluon Plasma?}
\author{Raghunath Sahoo\footnote{Email: Raghunath.Sahoo@cern.ch}}
\affiliation{Discipline of Physics, School of Basic Sciences, Indian Institute of Technology Indore, Simrol, Indore 453552, INDIA}

\begin{abstract}
The proton-proton ($pp$) collisions at the Large Hadron Collider (LHC), CERN, Switzerland has brought up new challenges and opportunities in understanding the experimental findings in contrast to the conventional lower energy $pp$ collisions. Usually $pp$ collisions are used as the baseline measurement at the GeV and TeV energies in order to understand a possible high density QCD medium formation in heavy-ion collisions. However, the TeV $pp$ collisions have created a new domain of research, where scientists have started observing  heavy-ion-like features (signatures) in high-multiplicity $pp$ collisions. This warrants a relook into TeV $pp$ collisions, if at all QGP-droplets are produced in such collisions. In this presentation, I discuss some of the new findings and concepts emerging out in $pp$ collisions at the LHC energies along with some of the new emergent phenomena in particle production.

\end{abstract}
\date{\today}
\maketitle

\section{Introduction}
\label{intro}
Relativistic heavy-ion collisions aim at producing the primordial matter in the laboratory, thermodynamics of which is governed by  partons - quarks and gluons, instead of the usual hadrons. The Universe at its infancy, was believed to be filled with such a deconfined matter of quarks and gluons. To recreate and characterize such a system in the laboratory, proton-proton ($pp$) collisions are usually used as baseline measurements assuming no such creation of partonic medium and deconfinement transition in these collisions. This is the case till the Relativistic Heavy Ion Collider (RHIC), Brookhaven National Laboratory, USA $pp$ collisions at a center-of-mass energy, $\sqrt{s}$ = 200 GeV. However, going from RHIC to the CERN Large Hadron Collider (LHC), where there is 35-65 times higher collision energies available in hadronic ($pp$) collisions, a new domain of particle production has possibly been created. This is because of new observations like strangeness enhancement, long-range correlations, collectivity in small systems ($pp$) etc., which are initially perceived to be signatures of Quark-Gluon Plasma (QGP) and expected to be seen in central heavy-ion collisions \cite{Heiselberg}. The LHC energy frontier opens up new challenges in understanding the multiplicity dependent experimental data and further characterizing the produced systems using differential variables.

A deconfined state of partons is expected either at very high temperatures and energy densities pertaining to an early universe scenario (LHC) or through the compression of nuclei so that the physical boundaries of the hadrons vanish so as to create a
free domain in which the partons seem to roam around, the volume of which is higher than the hadronic volume. The latter is expected to happen at lower collision energies, possibly at the FAIR or NICA. When the former points to a net-baryon free regime, the latter is baryon-rich domain of the Quantum Chromodynamics (QCD) phase space. For the present discussions, we shall stay close to the early universe scenario, which is the LHC energy domain. The chemical freeze-out temperature where all the inelastic particle producing interactions stop and the initial energy density measured at the LHC energies are much higher than the lattice QCD prediction for a deconfinement transition. These are the two basic requirements after which, one looks for all the possible indirect signatures 
of QGP, as it is highly short-lived. The kinetic freeze-out temperature, $T_{\rm kin}$ measured from the simultaneous Blast-wave model fitting to the low-$p_T$ multi-strange particle data taken by the ALICE experiment in $pp$ collisions at $\sqrt{s}$ = 7 TeV reveals that $T_{\rm kin}$ is comparable to that one obtains in p-Pb and Pb-Pb collisions. This is found to be higher than the critical temperature ($T_{\rm c}$) for the deconfinement transition \cite{ALICE-Nature}. 

\section{Basic Requirements for the formation of QGP}
It is well-known that high energy accelerators help us to revisit the earlier and hotter history of our Universe searching for a new simplicity by observing phenomena and particles no longer observed in our everyday experience. In this regards, three small and fundamental equations of Physics play a crucial role. Those are the famous de Broglie equation:  $E \propto 1/size$, Einstein's equation: $E = (\sum_i m_i)c^2$ and Boltzmann equation: $E= kT$. Higher collision energies thus help in probing lower length scales, producing a spectrum of new particles and also creating very high temperatrues, which pertain to an early time scenario. Through the head-on collisions of heavy nuclei like Au+Au or Pb+Pb one creates very high temperature (almost $10^5$ times the core of the Sun) and energy density (order of magnitude higher than the normal nuclear matter density) in the laboratory. Our theoretical estimations driven by lattice QCD for a deconfinement transition to happen are the critical temperature, $T_c \sim 150-170$ MeV and the critical energy density, $\epsilon_c \sim ~1 \rm ~GeV/fm^3$ \cite {lattice,lattice1}.
It is also seen that near this criticality the degrees of freedom of the system rises sharply \cite{lattice3}. The new deconfined phase of partons is highly short-lived (lifetime $\sim 10^{-23}$ sec) and hence the signatures of such a plasma are all indirect in nature. The major signatures of QGP are: strangeness enhancement, elliptic flow (and the higher order harmonics), suppression of quarkonia, long-range correlations, higher degree of collective radial expansion etc. QGP is perceived to be a thermally equilibrated (local) state of partons in a domain of volume, which is higher than the hadronic dimension. Thermalization is a requirement for the QGP formation and this can happen through mutual interactions of the system quanta given the volume of the fireball is sufficient enough or in a smaller volume like that of $pp$ collisions, this could possibly also happen through multiple interactions leading to the phenomenon of multipartonic interactions (MPI), if the quanta have sufficient momentum. This also goes inline with the requirement of a hydrodynamic scenario in $pp$ collisions, where one expects the mean free path of the system to be less than the system size and the space-time variation of local thermodynamic quantities should be less than the thermalization rate of the system, which is controlled by microscopic interactions. Although it was expected from the start of the RHIC at BNL that QGP will behave like an ideal gas and approach the Stefan-Boltzmann limit of different thermodynamic observables, after the discoveries are made, it was revealed that it behaves like a ``{\it perfect fluid}", with minimum shear viscosity to entropy ratio, $\eta/s$ seen in nature \cite{STAR-White}. Although QGP was discovered at the RHIC energies, the LHC heavy-ion runs were expected to bring new domains of particle production and some of the QGP signatures would depend on collision energy, like that was seen for the degree of quarkonia suppression at the LHC, which was found to be smaller than that of RHIC. The LHC $pp$ collisions, however brought up new scaling laws in the final state irrespective of the collision energy and collision species, showing the final state multiplicity is the driving (scaling) observable. Let's now look into some of the important observations in LHC $pp$ collisions before drawing any conclusion or giving any outlook for the possible formation of QGP-droplets in such hadronic collisions at the LHC \cite{Sahoo:2019ifs}.

\begin{figure}[ht!]
\includegraphics[scale=0.5]{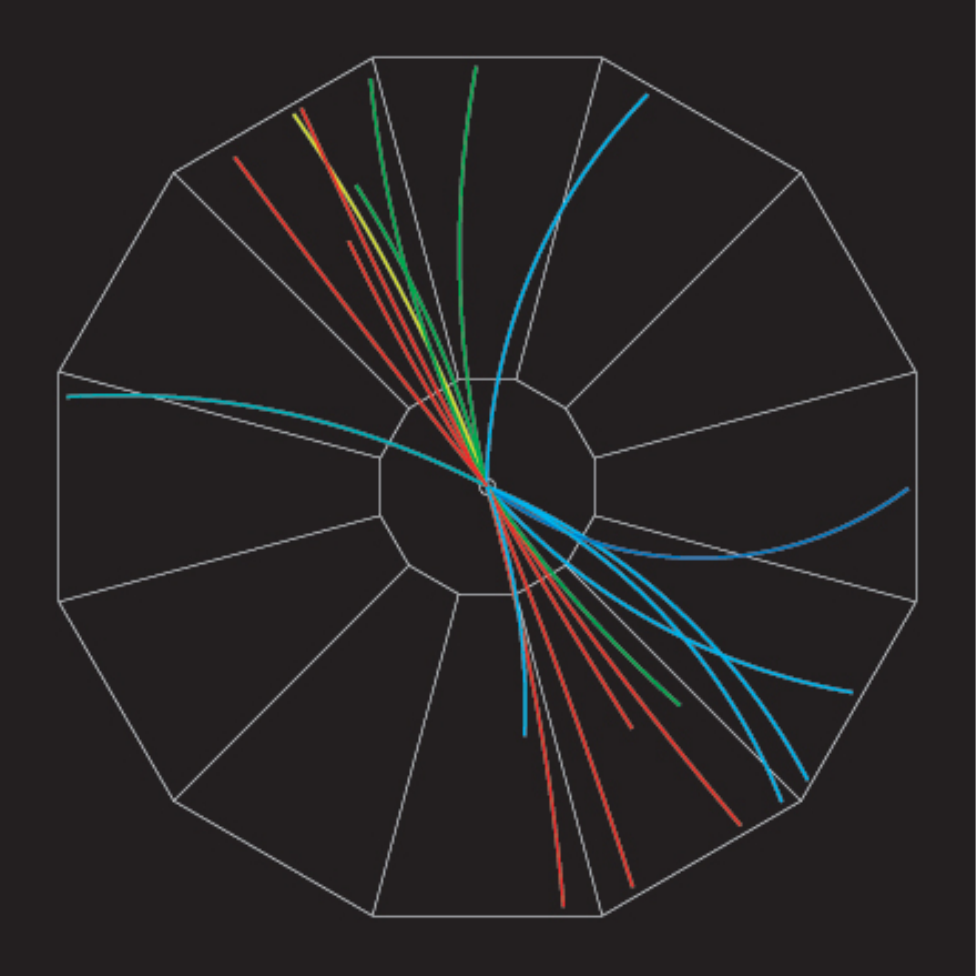}
\caption[]{(Color Online) Event topology of $pp$ collisions at $\sqrt{s}$ = 200 GeV in RHIC. The event shows a back-to-back jet structure. Figure courtesy:  STAR Experiment@RHIC \cite{star}.}
\label{fig1}
\end{figure}

\begin{figure}[ht!]
\includegraphics[scale=0.45]{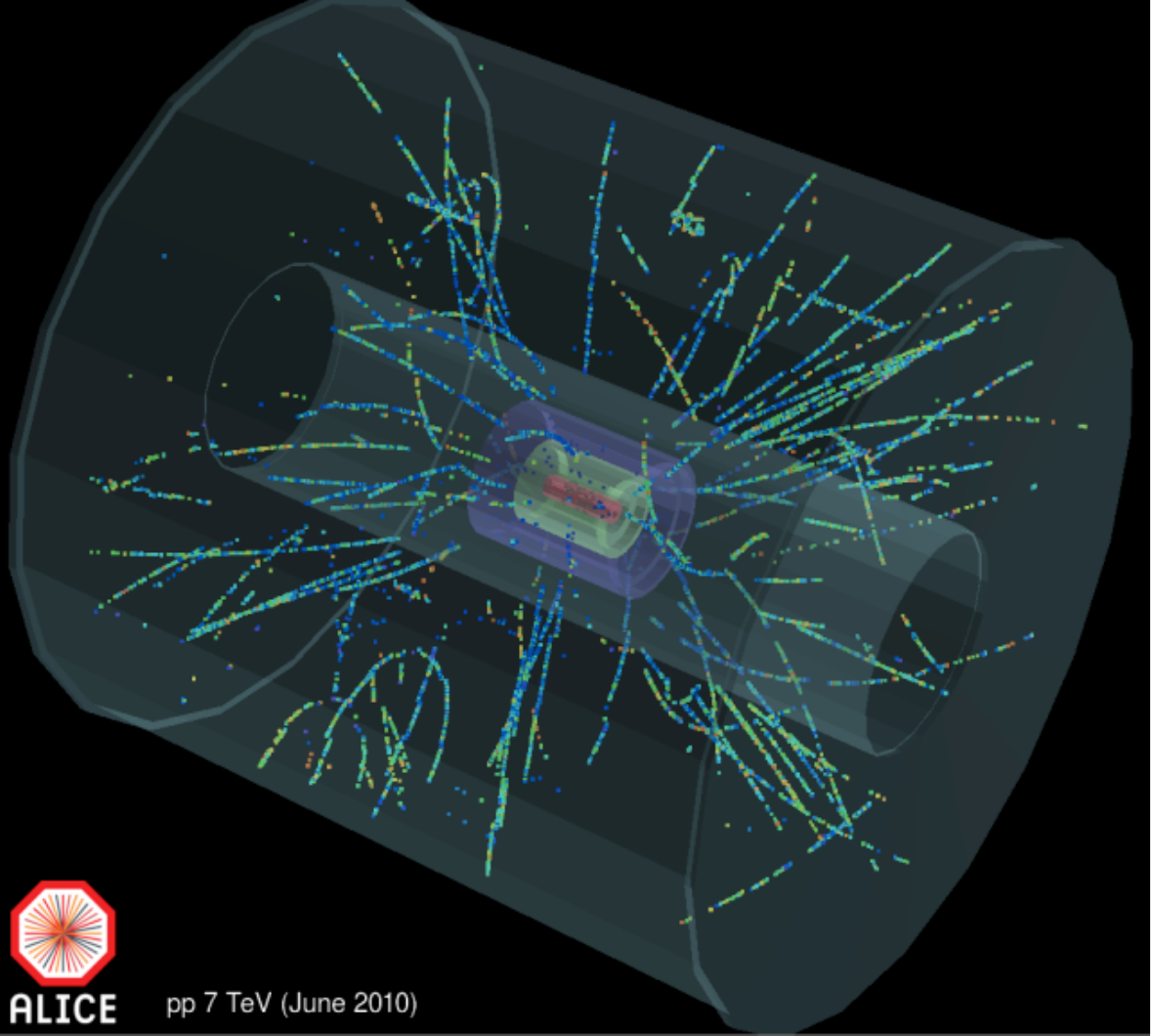}
\includegraphics[scale=0.45]{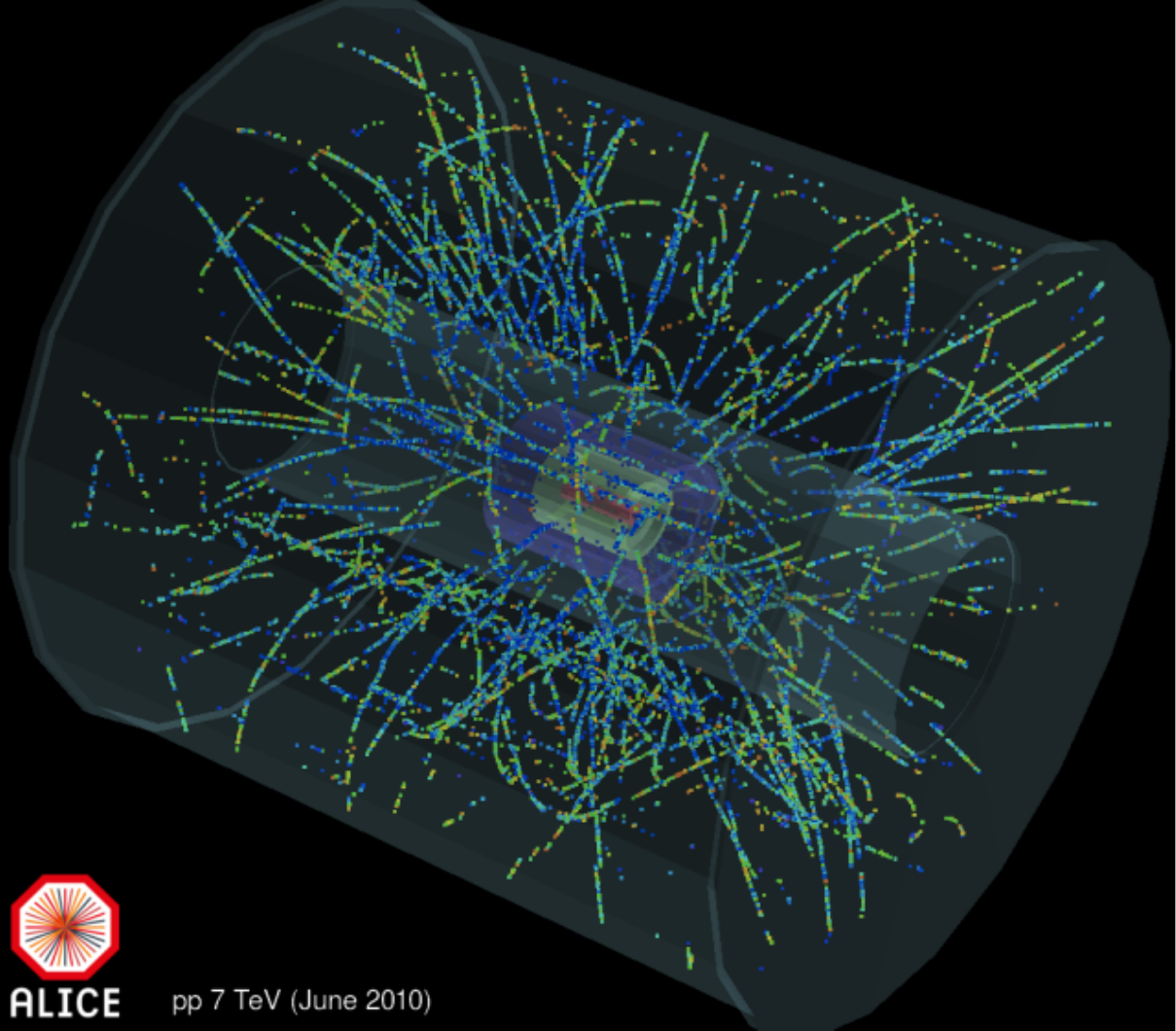}
\includegraphics[scale=0.45]{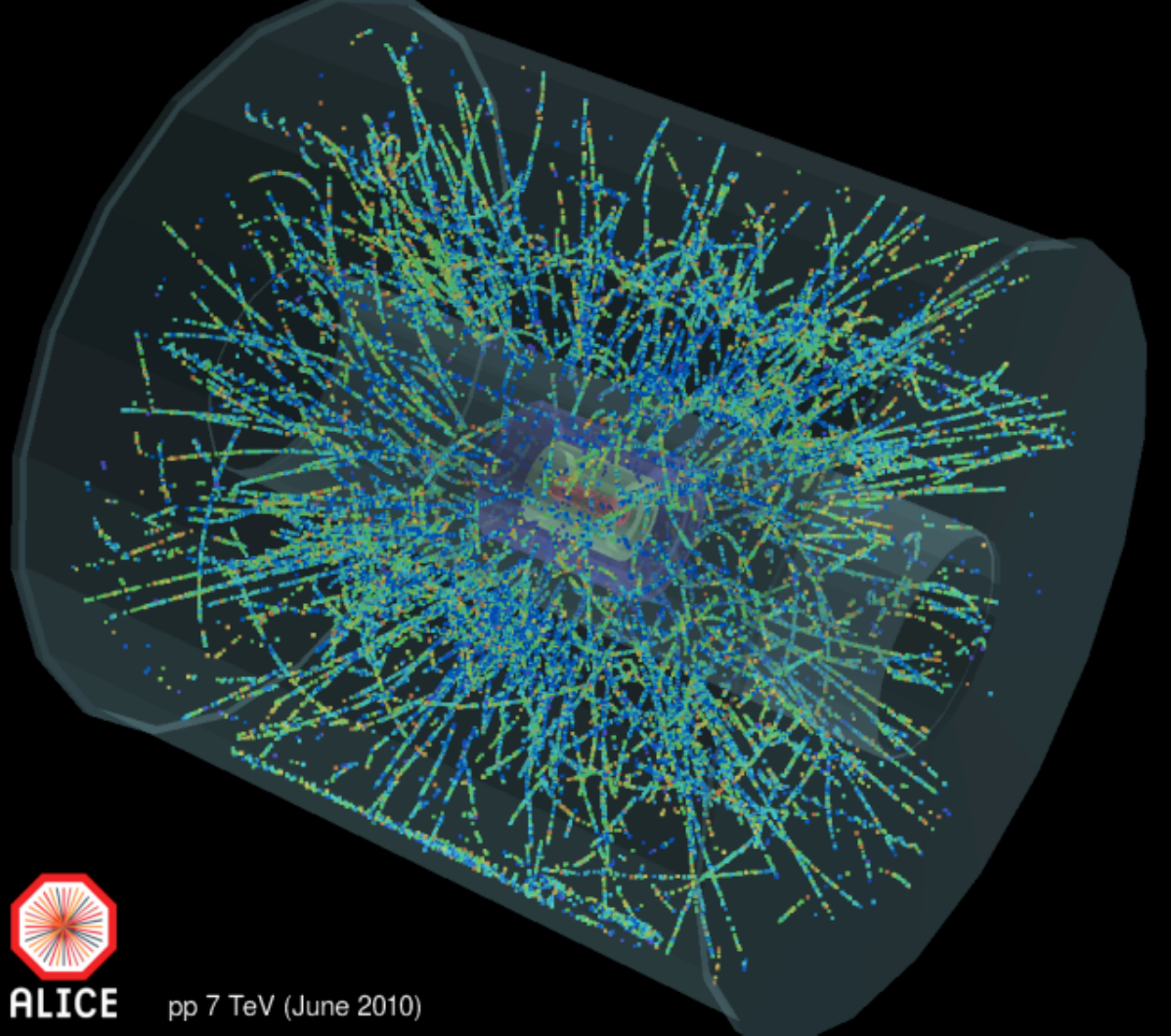}
\includegraphics[scale=0.45]{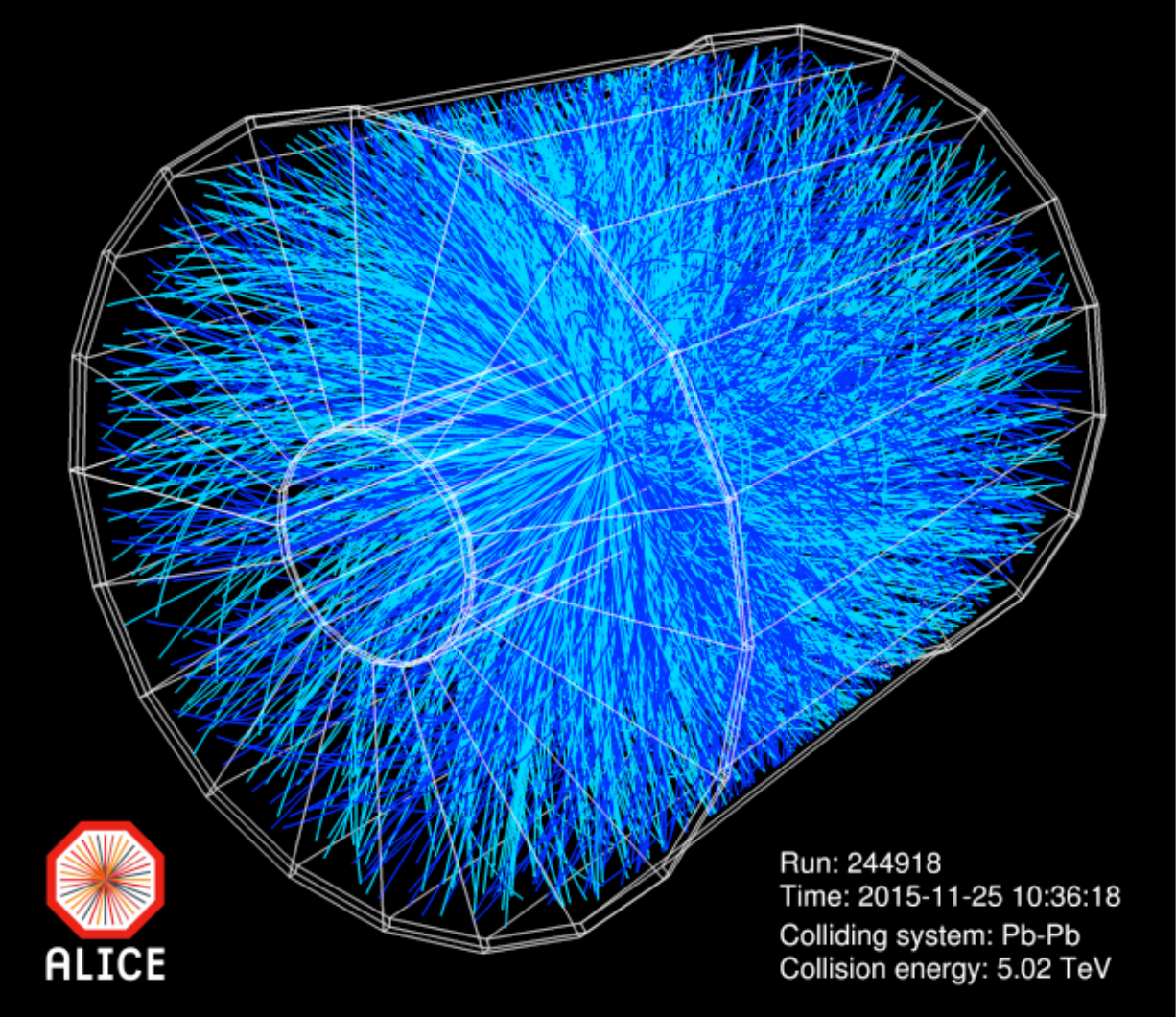}
\caption[]{(Color Online) Event topology of $pp$ collisions at the LHC energies showing the evolution of particle density in phase space: top to bottom - low-multiplicity, midium-multiplicity and high-multiplicity $pp$ collisions compared with heavy-ion collisions at the TeV energies. Figure courtesy: ALICE Experiment@CERN \cite{CDS}.}
\label{fig2}
\end{figure}

\section{Proton-Proton Collisions at the LHC}
The ALICE experiment has observed the enhancement of multi-strange particles in high-multiplicity $pp$ collisions and the enhancement is found to be comparable with p-Pb and Pb-Pb collisions at similar multiplicities \cite{ALICE-Nature}.
A simultaneous fitting of Blast-wave model to the low-$p_T$ (bulk) of the spectra of multi-strange particles shows a degree of collective radial flow ($<\beta> = 0.49 \pm 0.02$) and kinetic freeze-out temperature ($T_{\rm kin}$ = 163 $\pm$ 10 MeV), which is comparable to the critical temperature, $T_c$, while hardening of the spectra is also seen towards higher multiplicity classes \cite{ALICE-Nature}. The CMS experiment has parallely come up with a highly interesting observation of near-side long-range ridge-like two particle correlations in high-multiplicity $pp$ collisions \cite{Khachatryan:2010gv}, which was first observed in heavy-ion collisions at the RHIC energies.
  There are attempts to understand these heavy-ion-like observations in LHC $pp$ high-multiplicity events in order to conclude if these indicate to a new physics or QGP-droplets are formed in these hadronic collisions. In view of this, there are explorations through different theoretical models to confront to these experimental observations. 
  
  The strangeness enhancement in high-multiplicity $pp$ collisions is explained by theoretical models like DIPSY,
where the interactions between gluonic strings are allowed to form color ropes and through the mechanism
of “rope hadronization”, strangeness is produced. In addition, the radial flow observed in $pp$ collisions is also explained through ``color reconnection" mechanism in PYTHIA8. In the direction of small systems and thermalization, which is always a matter of great debate, if one assumes the availability of sufficiently energy leading to the participating partons to have enough momentum to go through multiple interactions, the emerging phenomenon, MPI happens to explain some of the features observed in experiments. For instance MPI seems to explain the forward rapidity J/$\psi$ production in $pp$ collisions at the LHC \cite{Thakur:2017kpv}. Although long-range near-side ridge in two particle correlation has a hydrodynamic origin, small system hydrodynamics is a question by itself. As explained above, one can expect hydrodynamic collectivity at the partonic level possibly through multipartonic interactions and this could be a reality at TeV energies, when one starts to see the dominance of gluons and sea quarks in the system. As explained in Ref. \cite{Bjorken}, the observed ridge-like structure in high-multiplicity $pp$ events by the CMS experiment  could be the evidence for the collision of aligned high-intensity flux tubes connecting the valence quarks of the colliding protons. The possible formation of a high density QCD medium however warrants the observation of elliptic flow and possibly higher order flow harmonics in $pp$ collisions. In addition, we also observe the inelastic $pp$ cross section, $\sigma_{pp}$ to show a different functional behavior at the LHC energies compared to the top RHIC energy \cite{Antchev:2017dia}. This possibly invokes a new domain of particle production which not only involves hard-QCD processes at the partonic level, but also phenomena like initial and final state radiation (ISR, FSR), underlying events (UE), color reconnection (CR), multipartonic interactions (MPI), different hadronization phenomena like rope hadronization.
  
\section{Event Topology Studies in Proton-Proton Collisions}
The event topology of proton-proton collisions at GeV energies is usually a back-to-back emission of momentum conserving dijet structure as shown in Fig. 1. However, at the LHC energies the event multiplicity in $pp$ collisions grows towards higher track densities in the phase space, as shown in Fig. 2. The underlying physics of these high-multiplicity events are yet to be fully understood. At the meantime, as we know particle production has soft (low-$p_T$ transfer) and hard components (dominated by pQCD high $p_T$ processes), one needs to devise new observables to study the event topology and thus disentangle event types to make deeper studies. In this direction, transverse spherocity ($S_0$) and $R_T$ (a self-normalized measurement of number of charged tracks in the transverse region: $N_T/<N_T>$ \cite{Martin:2016igp}) are some of the new event shape observables. Transverse spherocity is defined as:

\begin{eqnarray}
S^{pherocity}_{T}=S_{0}=\frac{\pi^{2}}{4} \min\limits_{ \vec{n}=(n_{x},n_{y},0)} \left( \frac{\sum_{i} |\vec{p}_{Ti}\times \vec{n} |}{\sum_{i} | \vec{p}_{Ti}|} \right)^{2},
\end{eqnarray}
where the values of $S_0$ which defines the structure in the transverse plane runs from 0 (for jetty) to 1 (for isotropic). In spherocity there is only $\phi$ dependence and only back-to-back dijet shapes can be selected.
\begin{figure}[ht!]
\includegraphics[width=2.4in]{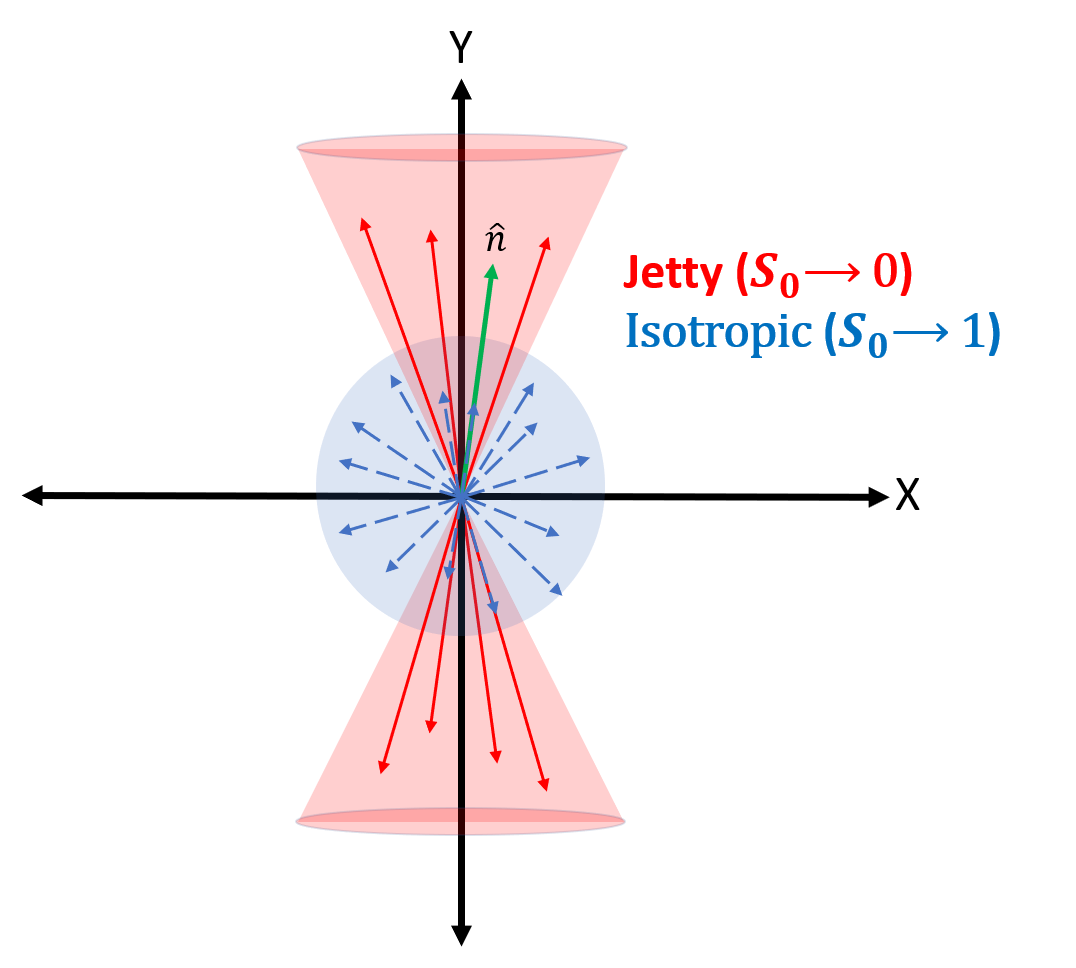}
\includegraphics[width=2.3in]{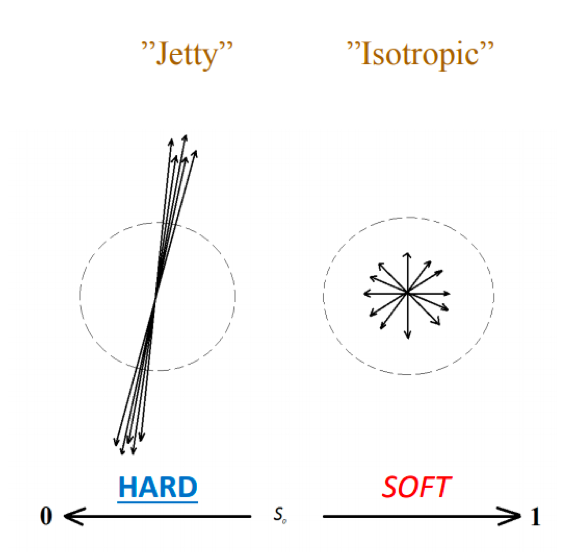} 
\caption[]{(Color Online) A schematic of event topology of $pp$ collisions showing jetty and isotropic events.}
\label{fig3}
\end{figure}
A schematic picture showing the selection of event topology using spherocity in $pp$ collisions is shown in Fig. 3.
The spherocity distribution of minimum bias events as a function of charged particle multiplicity in $pp$ collisions at $\sqrt{s}$ = 5.02 \cite{Khatun:2019dml} is shown in Fig. \ref{fig4} (top panel). One observes here that the high-multiplicity events have a higher degree of having isotropic event topology. The bottom panel of the figure shows isotropic events are dominated by soft QCD processes, whereas the jetty topology is hard QCD driven. These are obtained from PYTHIA8 simulated events \cite{Khatun:2019dml}. Moving from $pp$ to p+Pb and Pb+Pb collisions, it becomes evident to presume a smooth evolution of final state multiplicity and also the process of isotropization, given that more and more particles are produced as the system size grows.
Event shape studies are highly helpful in separating the event types and thereby the underlying particle production dynamics. Differential studies using final state multiplicity and event topology would be very useful in understanding the $pp$ physics at the LHC energies \cite{Cuautle:2015kra}.

\begin{figure}[ht!]
\includegraphics[scale=0.48]{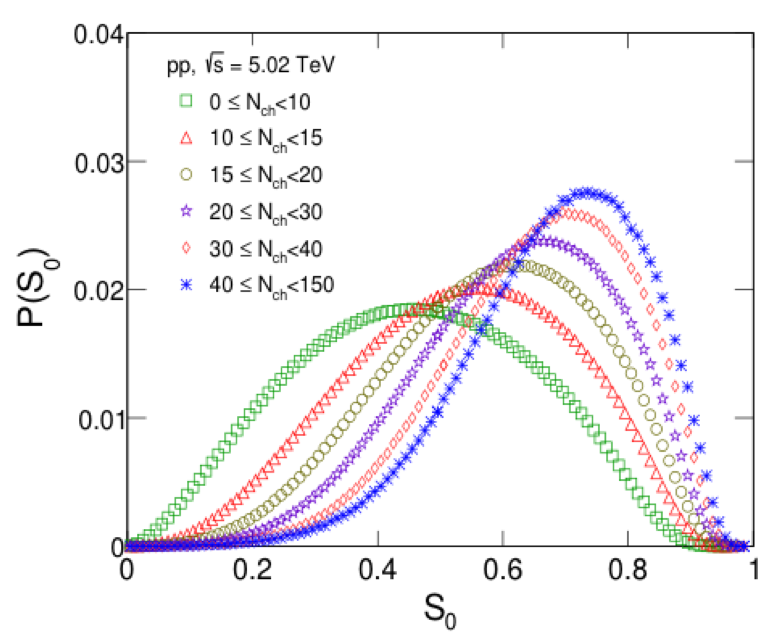}
\includegraphics[scale=0.50]{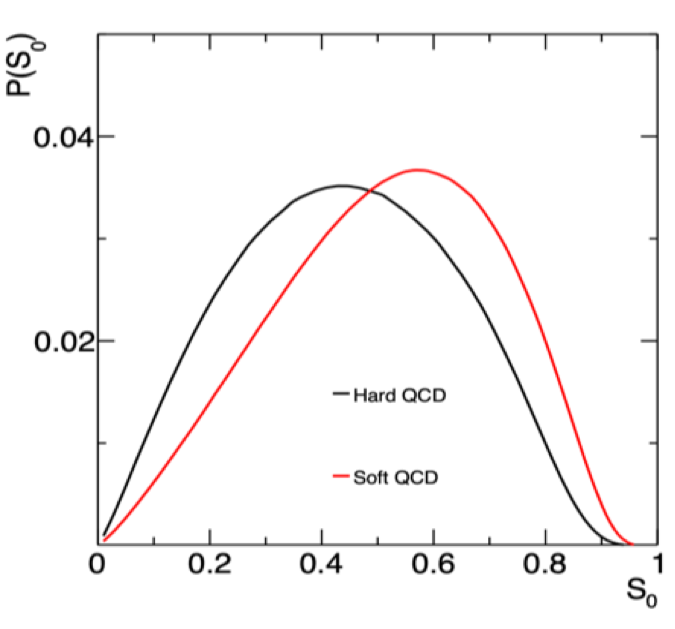}
\caption[]{(Color Online) (Top) The spherocity distribution of minimum bias events as a function of charged particle multiplicity in $pp$ collisions at $\sqrt{s}$ = 5.02 \cite{Khatun:2019dml}. High-multiplicity events have a higher degree of isotropic event topology. (Bottom) Isotropic events are dominated by soft QCD processes, whereas the jetty topology is hard QCD driven.}
\label{fig4}
\end{figure}

 \section{Some Emerging Phenomena in Proton-Proton physics}
  \label{emerging}
Some of the emergent physics aspects in view of the $pp$ collisions at the TeV energies are outlined below.
\begin{enumerate}
\item Particle multiplicity seems to drive the system properties. Threshold of $N_{\rm ch} \sim$ 20  for MPI to be dominant is observed, which is also seen as a thermodynamic limit where statistical ensembles give similar results \cite{Thakur:2017kpv,Natasha}. $p_{\rm T} \sim$ 8 GeV/c indicates a new domain of particle production \cite{Rath:2019cpe}.
\item Multipartonic interactions, Color reconnection, Rope Hadronization seem to explain some of the features observed in experiments.
\item Small system QGP-like behaviour is a future direction of research and needs more explorations both theoretically and through new observables in experiments to conclude about possible formation of QGP-droplets in $pp$ collisions at the TeV energies.
\item Connecting particle production from hadronic to nuclear collisions using various theoretical models involving  transport equations are necessary.
\item Application of non-extensive statistical mechanics in hadronic collisions is driven by particle spectra and this indicates to a new domain of studies on systems away from thermodynamic equilibrium.
\item QCD thermodynamics: $pp$ versus heavy-ion collisions and the multiplicity dependence of thermodynamic parameters.
\item Event topology studies using different observables like transverse spherocity, $R_T$ etc. and their experimental biases are some of the ongoing directions of research.
\end{enumerate}
Although this is a small list, we anticipate this to be populated with time, when a deeper understanding of underlying physics is made.


\section{Summary} 
\label{sec:sum}   
In this contribution, we have tried to summarize some of the important heavy-ion-like observations in $pp$ collisions at the LHC energies. The present theoretical understandings are also discussed, along with some outlook on the event topology and the idea of looking into high-multiplicity $pp$ events in order to explore the possibilities of formation of QGP-droplets at the LHC energies. This presentation leaves out with some of the open questions in $pp$ physics, which need a deeper understanding.
\section*{Acknowledgement}
The author would like to acknowledge the financial supports under the projects:  ALICE Project No. SR/MF/PS-
01/2014-IITI(G) of Department of Science \& Technology, Government of India and DAE-BRNS Project No. 58/14/29/2019-BRNS of Government of India.


{}

 \end{document}